\documentclass[10pt,conference]{IEEEtran} 
\IEEEoverridecommandlockouts
\usepackage{cite}
\usepackage{amsmath,amssymb,amsfonts}
\usepackage{algorithmic}
\usepackage{graphicx}
\usepackage{textcomp}
\usepackage{xcolor}
\def\BibTeX{{\rm B\kern-.05em{\sc i\kern-.025em b}\kern-.08em
    T\kern-.1667em\lower.7ex\hbox{E}\kern-.125emX}}

\usepackage{booktabs,arydshln}
\makeatletter
\def\adl@drawiv#1#2#3{%
        \hskip.5\tabcolsep
        \xleaders#3{#2.5\@tempdimb #1{1}#2.5\@tempdimb}%
                #2\z@ plus1fil minus1fil\relax
        \hskip.5\tabcolsep}
\newcommand{\cdashlinelr}[1]{%
  \noalign{\vskip\aboverulesep
           \global\let\@dashdrawstore\adl@draw
           \global\let\adl@draw\adl@drawiv}
  \cdashline{#1}
  \noalign{\global\let\adl@draw\@dashdrawstore
           \vskip\belowrulesep}}
\makeatother
\usepackage{multirow}

\makeatletter
\def\footnoterule{\kern-3\p@
  \hrule \@width 2in \kern 2.6\p@} % the \hrule is .4pt high
\makeatother
\usepackage{hyperref}

\begin{document}

\title{MusCaps: Generating Captions for Music Audio \\
\thanks{This work was jointly supported by UK Research and Innovation [grant number EP/S022694/1], Queen Mary University of London, and Universal Music Group.}
}

\author{\IEEEauthorblockN{Ilaria Manco\textsuperscript{*\dag}, Emmanouil Benetos\textsuperscript{*}, Elio Quinton\textsuperscript{\dag}
\& Gy\"orgy~Fazekas\textsuperscript{*}}
\IEEEauthorblockA{
\textit{\textsuperscript{*}School of EECS, Queen Mary University of London, U.K.} \\
\textit{\textsuperscript{\dag}Music \& Audio Machine Learning Lab, Universal Music Group, London, U.K.} \\ 
i.manco@qmul.ac.uk, emmanouil.benetos@qmul.ac.uk, elio.quinton@umusic.com, g.fazekas@qmul.ac.uk}
}

\maketitle

\begin{abstract}
Content-based music information retrieval has seen rapid progress with the adoption of deep learning. Current approaches to high-level music description typically make use of classification models, such as in auto-tagging or genre and mood classification. In this work, we propose to address music description via audio captioning, defined as the task of generating a natural language description of music audio content in a human-like manner. To this end, we present the first music audio captioning model, MusCaps, consisting of an encoder-decoder with temporal attention. Our method combines convolutional and recurrent neural network architectures to jointly process audio-text inputs through a multimodal encoder and leverages pre-training on audio data to obtain representations that effectively capture and summarise musical features in the input. Evaluation of the generated captions through automatic metrics shows that our method outperforms a baseline designed for non-music audio captioning. Through an ablation study, we unveil that this performance boost can be mainly attributed to pre-training of the audio encoder, while other design choices – modality fusion, decoding strategy and the use of attention -- contribute only marginally.
Our model represents a shift away from classification-based music description and combines tasks requiring both auditory and linguistic understanding to bridge the semantic gap in music information retrieval\footnote{Code available at \href{https://github.com/ilaria-manco/muscaps}{https://github.com/ilaria-manco/muscaps}}.
\end{abstract}

\section{Introduction}\label{sec:introduction}
\begin{figure*}[t]
\centering
\includegraphics[scale=0.4]{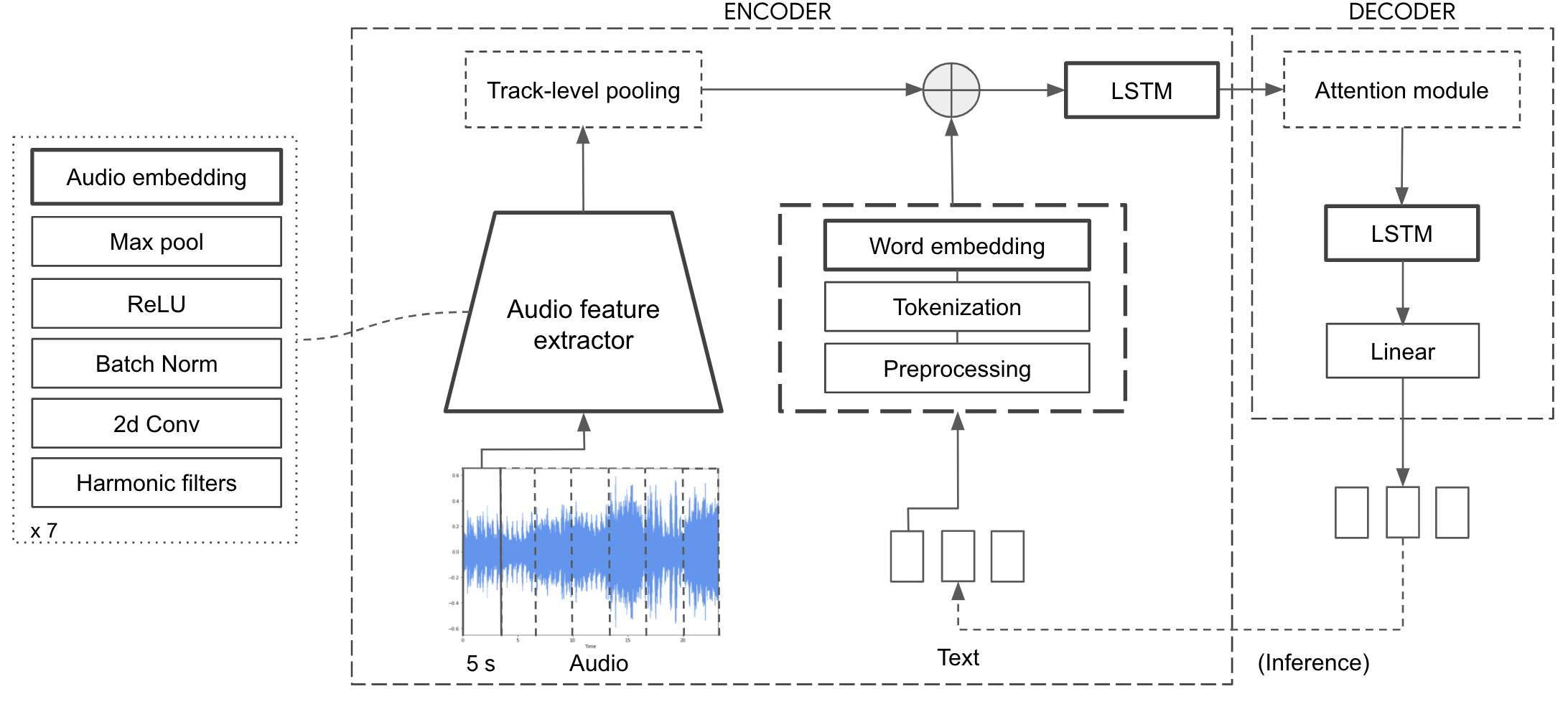}
\caption{Overview of the MusCaps architecture. Our model has a hybrid and modular design that allows us to join both convolutional and recurrent neural network architectures to process the audio-text input pairs, optionally learn a soft alignment between them and output a sequence of predicted text tokens conditioned on the input audio.}
\label{fig:archi}
\vspace{-0.5cm}
\end{figure*}

Current music information retrieval (MIR) approaches to music description typically rely on single- or multi-label classification. A prominent example is music auto-tagging \cite{Choi2016a, Kim2017, Pons2018}, in which descriptive keywords are assigned to a music clip so as to convey high-level characteristics of the input such as genre, instrumentation and emotion. While this offers a reasonable starting point to modelling high-level representations of music, its output is limited to a predefined set of categorical labels. This in turns limits its usefulness in applications such as music search and recommendation, which would benefit from being able to both process and generate more human-like, detailed and nuanced descriptions.
It is in fact through natural language that we often query music collections and search for known and unknown music content. A significant part of our daily language is also devoted to describing the musical world, offering an interface between music as mere audio signals and the set of abstractions used to describe it.
While a wealth of music information is encoded in text, research in machine listening and MIR has traditionally overlooked the relationship between audio and natural language.

In this study, we focus on the novel task of \textit{music captioning}, which we define as the ability to extract, disentangle and reason about high-level musical concepts in music audio, and then map these to the text modality by generating syntactically and semantically correct sequences of words. This is significantly harder than other, more common music description tasks such as classification and recognition, but presents a solution to some of their limitations. Captioning systems not only need to recognise signal-level features such as instrumentation and high-level descriptors such as genre, but they must also encode the relationship between them, thus better capturing the nuances of musical content; they also produce fully formed, descriptive sentences, that more closely match human queries. Through its joint use and processing of audio and linguistic information, music captioning also provides a first step towards the development of audio-and-language models for music understanding.

Finally, music captioning has several useful applications, such as producing descriptions for items in large music catalogues or vast collections of amateur and user-generated content; automatically generating evocative descriptions of music in films and videos for deaf and hard-of-hearing people; enabling search and discovery of music through more human-like queries; and providing explanations for automatic music recommendations. 

To the best of our knowledge, this is the first work on music audio captioning. In the absence of benchmark datasets and established research on this task, we build upon previous literature on image and audio captioning and compare our model performance to a baseline sequence-to-sequence model. Differently from pioneering work on neural architectures for audio captioning, typically composed of an audio encoder and a text decoder, we propose a multimodal encoder that learns a joint representation of both audio and text to better account for the need to capture high-level semantics and summarise information that emerges at different levels of granularity in the input. 
We also show that, while audio-text data in the music domain is hard to obtain, this data scarcity issue can be effectively alleviated in the context of music captioning by employing suitable pre-trained audio representations. To accomplish this, we leverage large-scale pre-training on a publicly available music dataset and investigate the role of this pre-training step when only a smaller corpus is available for the downstream task of music audio captioning. 

The main contributions of this work can be summarised as follows: (i) we propose the first music captioning model for track-level audio and evaluate it on the captioning and audio retrieval tasks; (ii) we establish whether commonly available pre-trained music auto-tagging models can be usefully employed in a transfer learning setting for downstream audio-linguistic tasks; (iii) through an ablation study, we investigate the effect of modality fusion, temporal attention and beam search decoding on the model performance.

\section{Related Work}\label{sec:related_work}
\subsection{Vision \& Language}
Within the field of machine perception, image caption generation has long been studied \cite{Vinyals2014, Xu2015a}. Thanks to enormous progress in natural language processing (NLP) and computer vision, coupled with the increased availability of large-scale image-text datasets, research in this field has recently focussed on developing models to solve several vision-and-language (V\&L) tasks. Most V\&L models can be classified under two paradigms: CNN-RNN \cite{Vinyals2014, Xu2015a, Karpathy2017} and Transformer-based \cite{Lu2019, Su2020, Haroldli2019, Zhou2020}. The former encompasses a family of fusion-based models in which convolutional neural network (CNN) architectures are employed as feature extractors, while recurrent neural networks (RNN), are employed as language models conditioned on the visual input. More recently, the success of BERT \cite{Devlin2019} has motivated research into its adaptation to V\&L tasks, allowing to achieve state-of-the-art (SOTA) performance in all of them, both for still image and video \cite{Korbar2020, Sun2019}.

\subsection{Audio \& Language}
While multimodal tasks have long been studied in the visual domain, audio-and-language research has only recently started to emerge, with the first audio captioning model proposed in \cite{Drossos2017}. Following a similar development to its visual counterpart, audio captioning has seen a rapid progress over the last few years \cite{Wu2019a, Cakr2020, Eren2020, Kim2019b, Tran2020, Ikawa2019, PerezCastanos2020, Koizumi2020a}, greatly encouraged by the recently introduced DCASE Challenge dedicated to the task\footnote{http://dcase.community/challenge2020/task-automatic-audio-captioning}. Most prior audio captioning methods make use of encoder-decoder models, frequently including sequence modelling modules, such as RNNs \cite{Ikawa2019} or variants like gated recurrent units (GRU) \cite{Drossos2017, Eren2020} and long short-term memory (LSTM) networks \cite{Kim2019b}, in their encoder to take care of the temporal structure of audio inputs. Most of these works also make use of attention mechanisms to align the audio and text modalities \cite{Drossos2017, Kim2019b, PerezCastanos2020, Wu2019a}. More recently, following the success of self-attention in V\&L models, a small body of work has also started exploring the use of Transformer-based models in audio captioning \cite{Koizumi2020, Tran2020}.

Our work is inspired by CNN-RNN architectures developed for image and audio captioning, but focusses on how such approaches can be extended to the music domain for the first time. The only prior works that attempt a similar goal can be found in \cite{Choi2016} and \cite{Cai2020}. However, the method proposed in \cite{Choi2016} fails to generate grammatically correct sentences, while \cite{Cai2020} simplifies the task by reframing it as the generation of a sequence of tags. Similarly, prior work on audio-text representation learning also makes use of tags \cite{Favory2020}, while we stress that our approach focusses on natural language.

\subsection{Transfer Learning for MIR Tasks}
Finally, one of the underlying ideas of our study is to leverage pre-training on music data to alleviate the issue of data scarcity in tasks such as captioning where parallel data across modalities is required.
Tasks such as music tagging have been shown to successfully extract salient characteristics and learn musically relevant concepts such as mood, genre, era, and emotional content \cite{Pons2018, Choi2016a}. This suggests that the data representations learnt by networks designed for these tasks can act as useful descriptors for more complex tasks that are highly dependent on similar properties of the input data. Previous work has demonstrated the benefit of using these pre-trained features in a transfer learning setting for regression and classification tasks in the music domain \cite{VanDenOord2014,Choi2017}, but this has not yet been explored in a multimodal setting. 

\section{Proposed Method}\label{sec:method}
Our model is composed of five main building blocks: a text embedding module, an audio feature extractor, a multimodal encoder, an attention mechanism and a natural language decoder. An overview of the architecture is presented in Fig. \ref{fig:archi}. In what follows, we offer a detailed description of each component.

\subsection{Text Embedding} \label{sec:sentence_emb}
The text input is tokenized and encoded through an embedding matrix of dimensions $V \times d$, where $V$ is the vocabulary size and $d$ is the word embedding dimension. This is initialised with 300-dimensional GloVe word embeddings \cite{Pennington2014}, pre-trained on Wikipedia 2014 and Gigaword 5 data, and kept frozen while training our model. Each sentence is thus encoded as a sequence $\mathcal{S} = \{\boldsymbol{w}_1, ..., \boldsymbol{w}_{T}\}, \boldsymbol{w}_t \in \mathbb{R}^d$, where $T$ is the variable length of the sentence.

\begin{figure}[t]
\centering
\includegraphics[scale=0.45]{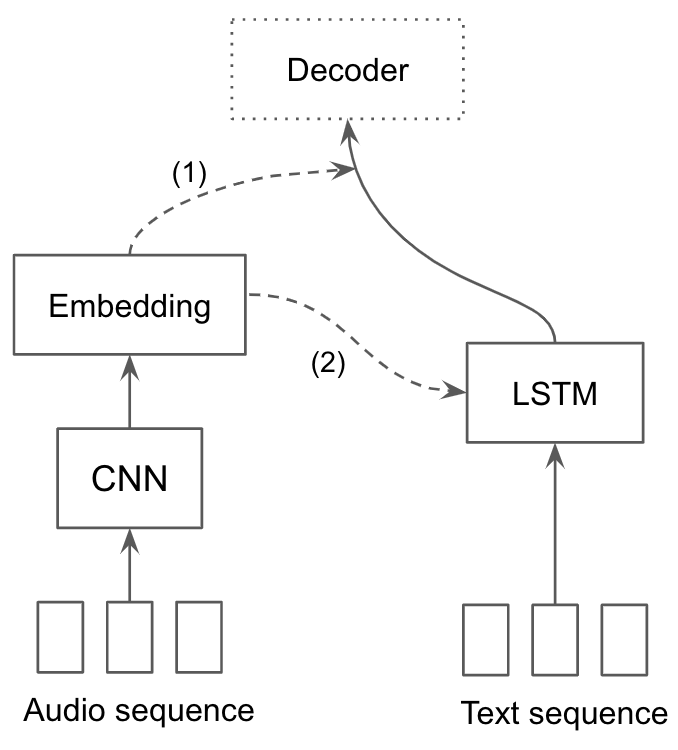}
\caption{Illustration of the two fusion strategies: (1) when using late fusion, the fixed-length representation of the whole input audio is only fused before being fed to the decoder; (2) in the early-fusion approach, the audio representation is concatenated to the word embedding and fed to the LSTM at each step.}
\label{fig:fusion}
\vspace{-0.5cm}
\end{figure}

\subsection{Audio Feature Extractor} \label{sec:audio_feature}
The audio feature extraction module is designed to take a variable-length raw audio input, split it into fixed-length chunks and extract features to be fed to the encoder. In summary, its role is to encode audio into a fixed-length representation or a sequence thereof. Inspired by the established practice of using CNN-based image feature extraction for vision-and-language tasks \cite{Lu2019,  Korbar2020, Haroldli2019} and further motivated by the development of similar feature extraction networks specifically designed for music audio data \cite{Pons2018, Choi2017}, we make use of convolutional neural networks for music auto-tagging as our feature extractors.

Among several architectures, we select two in our study, based on their performance on the music tagging task and their musically informed design: \textit{Musicnn} \cite{Pons2018}, which combines vertical and horizontal convolutional filters to capture both timbral and temporal features, and \textit{Harmonic CNN} \cite{Won2020a}, which uses trainable filters to model harmonic structures in the audio. A comprehensive comparison of these two networks, among other architectures, is presented in \cite{Won2020}. We discard the classification layers of the networks, using only the front-end and convolutional layers in our feature extraction, and use pre-trained weights\footnote{https://github.com/minzwon/sota-music-tagging-models} from \cite{Won2020}.

For each $n$-second chunk, where $n$ is the input length of the feature extraction network (3s for \textit{musicnn} and 5s for \textit{Harmonic CNN}, both at a sampling rate of 16 kHz), we obtain a feature vector of dimension $k$, such that the whole input sequence is encoded by a variably sized set of $L$ audio features $A = \{\boldsymbol{a}_1, ..., \boldsymbol{a}_{L}\}, \boldsymbol{a}_i \in \mathbb{R}^k$. When no attention mechanism (Section \ref{sec:attention_encoder}) is included in the model, we then obtain track-level features by applying average pooling on the feature maps across the time dimension. This yields $\tilde{\boldsymbol{a}} \in \mathbb{R}^k$, a track-level vector representation of the input audio.

\subsection{Multimodal Encoder} \label{sec:encoder}
In its baseline variant, our multimodal encoder consists of a 1-layer unidirectional LSTM. Using the index $t = 1, ..., T$ to denote the $t$-th word in a caption, the $t$-th hidden state of the encoder LSTM is given by 
\begin{equation}
    \boldsymbol{h}^{enc}_{t}=\operatorname{LSTM}\left(\boldsymbol{x}^{enc}_{t}, \boldsymbol{h}^{enc}_{t-1},  \boldsymbol{m}_{t-1}\right),
\end{equation}
where $\boldsymbol{x}^{enc}_{t}$ is the input and $\boldsymbol{m}_{t}$ is the cell state at step $t$.

In order to investigate the effect of using a multimodal input, we choose to perform modality fusion at different stages of the encoding procedure. In the \textit{early fusion} approach, embeddings of the audio-text input pair are concatenated and passed to the LSTM as an input at each step:
\begin{equation}
    \boldsymbol{x}^{enc}_{t} = \left[\boldsymbol{a}, \boldsymbol{w}_{t}\right],
\end{equation}
where $\boldsymbol{a}$ is obtained by passing the track-level audio embedding $\tilde{\boldsymbol{a}}$ generated as described in the previous section through a fully connected layer.
In the \textit{late fusion} approach, the encoder LSTM only takes the text as input ($\boldsymbol{x}^{enc}_{t} = \boldsymbol{w}_t$) and its hidden state $\boldsymbol{h}^{enc}_{t}$ is then concatenated with $\boldsymbol{a}$ only prior to being passed to the decoder:

\begin{equation}\label{eq:dec_input_late}
    \boldsymbol{x}^{dec}_t = \left[\boldsymbol{a}, \boldsymbol{h}^{enc}_{t-1} \right].
\end{equation}
In this case, similarly to what was done for early fusion, $\boldsymbol{a}$ is obtained by passing the audio embeddings through an additional linear layer, matching the dimension of the encoder hidden state ($\boldsymbol{W}_a \tilde{\boldsymbol{a}} \in \mathbb{R}^{H_{enc}}$).
As illustrated in Fig. \ref{fig:fusion}, in the early-fusion approach the audio content is therefore available to the encoder LSTM, while in the late-fusion approach the audio input does not influence the sequence dynamics modelled by the encoder. 

\begin{figure}[t]
\centering
\includegraphics[scale=0.4]{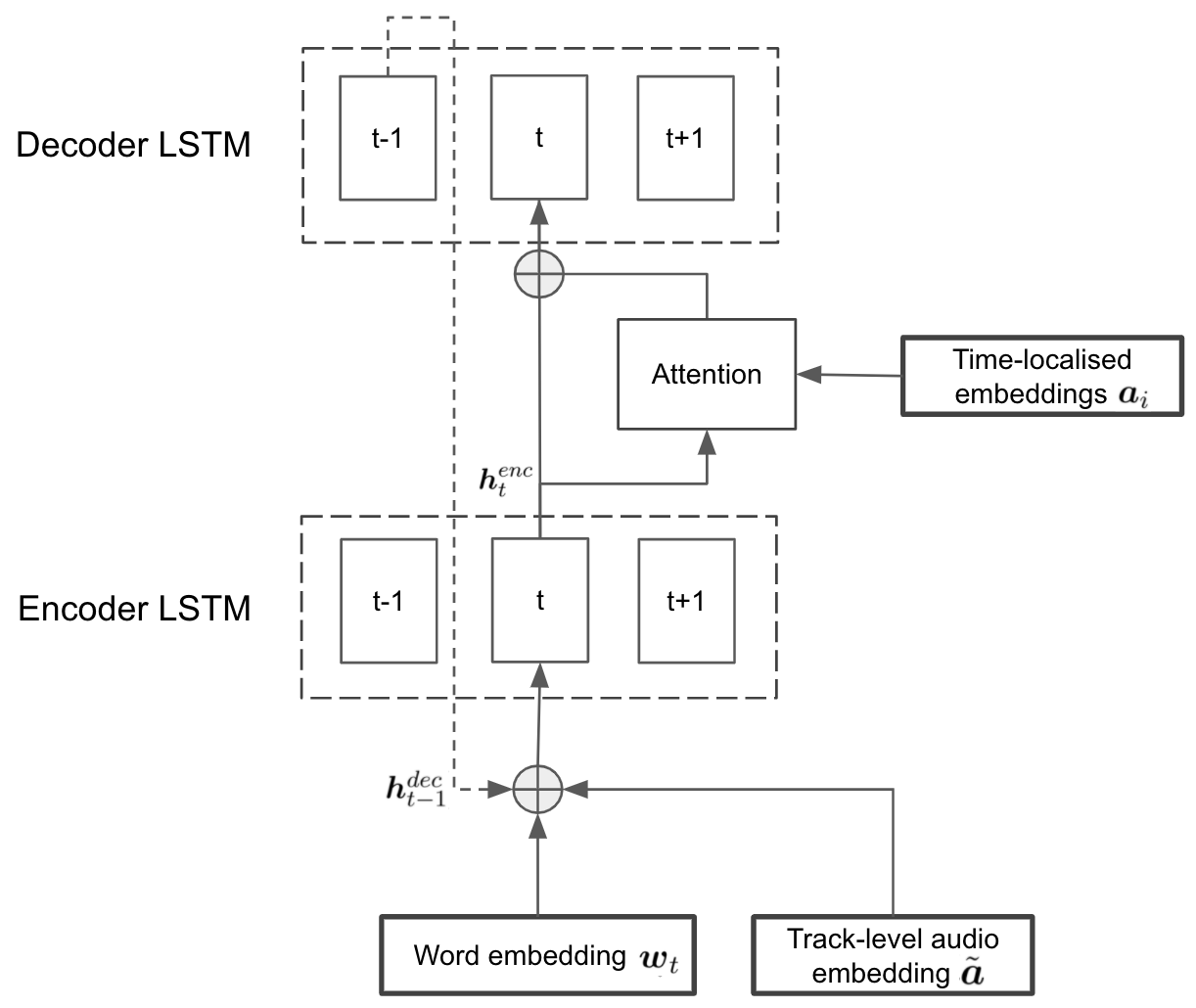}
\caption{Overview of the encoder-decoder architecture with the soft attention mechanism.  This is similar to the top-down attention in \cite{Anderson2017}.}
\label{fig:attention}
\vspace{-0.5cm}
\end{figure}

\subsection{Attention Mechanism} \label{sec:attention_encoder}
A key challenge in our multimodal encoder is to appropriately \textit{align} and \textit{summarise} the elements in the input audio to which an associated textual description is most sensitive to. We reduce this problem to tackling two distinct aspects: identifying the salient acoustic characteristics emerging from a global description of the input track and localising the temporal segments that are most strongly responsible for the associated description. An optimal approach would provide both and we note that the encoder described in the previous section may fall short of these requirements.
In light of the shortcomings of compressing the whole audio input into a static representation, we adopt a soft attention mechanism to dynamically weigh different temporal sections in the audio input. For this, we obtain time-localised audio embeddings $\boldsymbol{a}_i$ by using the audio features produced by the audio encoder over $n$-second segments prior to average pooling, as described in Section \ref{sec:audio_feature}.

The attention model used is a soft top-down attention mechanism, similar to approaches first proposed for the image captioning task in \cite{Xu2015a, Anderson2017}. This mechanism allows the decoder to attend to different sections of the input audio according to attention weights $\beta_{ti}$, which are based on an alignment score $e_{t i}$ learned from both the audio features $\boldsymbol{a}_i$ and the hidden state of the encoder $\boldsymbol{h}_{t}^{enc}$:

\begin{equation}\label{eq:attention}
    \begin{aligned}
    e_{t i} &= \boldsymbol{w}_{att}^{\intercal} \tanh \left(\boldsymbol{W}^{att}_{a} \boldsymbol{a}_{i}+\boldsymbol{W}^{att}_{h} \boldsymbol{h}_{t}^{enc}\right) \\
    \beta_{t i} &= \frac{\exp \left(e_{t i}\right)}{\sum_{k=1}^{N} \exp \left(e_{t k }\right)},
    \end{aligned}
\end{equation}
where $\boldsymbol{w}_{att} \in \mathbb{R}^{H_{dec}}$, $\boldsymbol{W}^{att}_{a} \in \mathbb{R}^{H_{dec} \times k}$,  $\boldsymbol{W}^{att}_{h} \in \mathbb{R}^{H_{dec} \times H_{enc}}$ are parameters to be learnt and $H_{enc}$ and $H_{dec}$ are the number of hidden units in the encoder and decoder LSTM.

From (\ref{eq:attention}), we obtain the attended vector $\hat{\boldsymbol{a}}_{t}$ as a weighted sum over the $L$ $n$-second chunks forming the audio input:

\begin{equation}
    \hat{\boldsymbol{a}}_{t}=\sum_{i=1}^{L} \beta_{t i} \boldsymbol{a}_{i}.
\end{equation}

\subsection{Language Model Decoder} \label{sec:decoder}
The input to the decoder LSTM varies depending on the fusion approach used at the encoding stage and whether the attention module is used. In the case of early fusion, it simply consists of the hidden representations produced by the encoder LSTM, while in the late-fusion case it is given by the audio features concatenated with the hidden states as in \eqref{eq:dec_input_late}.

Finally, if the attention module is added to the model architecture, the attended audio feature replaces the mean-pooled vector in \eqref{eq:dec_input_late}:
\begin{equation}
    \boldsymbol{x}^{dec}_t = \left[\hat{\boldsymbol{a}}_{t}, \boldsymbol{h}^{enc}_{t-1} \right].
\end{equation}

The overall architecture of the LSTM encoder, attention module and decoder is illustrated in Fig. \ref{fig:attention}.

A linear layer, with parameters $\boldsymbol{W}_d$ and biases and $\boldsymbol{b}_d$, is then appended to the recurrent layers, followed by a softmax nonlinearity, thus acting as a classifier, where each class represents one of the words in the vocabulary. Its output is $Y = \{\boldsymbol{y}_1, ..., \boldsymbol{y}_{L}\}, \boldsymbol{y}_t \in \mathbb{R}^{V}$, a sequence of vectors representing the probability distribution over the word vocabulary at each step $t$ in a $T$-long sequence: 
\begin{equation}
P(\boldsymbol{y}_t | \boldsymbol{y}_{t-1}) = \operatorname{softmax} (\boldsymbol{W}_d \boldsymbol{h}^{dec}_{t} + \boldsymbol{b}_d),
\end{equation}
such that the joint distribution 
\begin{equation}\label{eq:joint}
P(Y  | A) = \prod_{t=1}^{T} P(\boldsymbol{y}_t | \boldsymbol{y}_{t-1})
\end{equation}
gives the probability distribution over full text sequences $Y$, given an audio input $A$.

During training, a caption is generated through greedy decoding by selecting the word with the highest probability at step $t = \{1, ..., L\}$. At inference time, beam search decoding is also explored as a post-processing step. In this case, a fixed number, equal to the beam size $b$, of most likely hypotheses up to step $t$ is considered to generate the output at $t+1$.

\begin{table*}[!t] 
\caption{Comparison with the baseline, evaluated on our test set. Both variants of MusCaps, each using a different pre-trained audio feature extractor, perform significantly better than the baseline model across all metrics.}
\vspace{-0.2cm}
\label{tab:results}
\centering
\begin{sc}
\begin{tabular}{lccccccccc}
\toprule
Model & BLEU\textsubscript{1} & BLEU\textsubscript{2} &  BLEU\textsubscript{3} &  BLEU\textsubscript{4} & METEOR & ROUGE\textsubscript{L} & CIDEr & SPICE & SPIDEr \\
\midrule
DCASE Baseline & $13.8$  & $5.8$  & $3.6$ & $2.0$ & $4.3$ & $13.7$ & $19.6$ & $5.6$ & $12.6$  \\
\cdashlinelr{1-10}
MusCaps-Musicnn &  $34.3$ & $15.0$ & $8.5$ & $5.4$ & $29.3$ & $39.9$ & $33.0$ &	$\mathbf{24.2}$ & $28.6$ \\
MusCaps-Hcnn & $\mathbf{37.3}$ & $\mathbf{16.4}$ & $\mathbf{9.3}$ & $\mathbf{5.9}$ & $\mathbf{29.6}$ & $\mathbf{40.6}$	& $\mathbf{36.9}$ & $23.5$ & $\mathbf{30.2}$  \\
\bottomrule
\end{tabular}
\end{sc}
\vspace{-0.5cm}
\end{table*}

\section{Experiments}\label{sec:experiments}
\subsection{Dataset}
In our experiments, we use a private production music\footnote{\emph{Production music} is written and recorded with the aim of being licensed for synchronisation in audio or audiovisual productions such as films and adverts. It is typically organised in catalogues and provided with metadata and descriptions to facilitate discovery of suitable content.} dataset consisting of audio-caption pairs. We denote such a dataset $\mathcal{D}= \{A_i, c_i\}^{P}_{i=1}$, where $P$ is the number of pairs, $A_i$ an audio track and $c_i$ the corresponding caption. We filter out unsuitable examples by  (i) keeping only tracks of length between 30 and 360 seconds; (ii) keeping only captions that contain between 3 and 22 tokens and that are not duplicated across the dataset. Through this cleaning procedure we retain $P = 6,035$ pairs. A random split was used to obtain training, validation and test sets with a 60/20/20 ratio.

The dataset is pre-processed by applying tokenization and encoding each token as a numerical ID. 
Special tokens, used for infrequent words (\textsc{$\langle unk \rangle$}), padding (\textsc{$\langle pad \rangle$}), start (\textsc{$\langle sos \rangle$}) and end of sentence (\textsc{$\langle eos \rangle$}) are then added to the vocabulary.

\subsection{Experimental Setting}\label{sec:settings}
Our caption generation model is trained using \textit{teacher forcing} \cite{Williams1989}. At each step $t$, the $t-1$ element of the target caption is supplied as text input during training, while at test time this is replaced by the previous decoder output. In practice this is achieved by providing the special start token \textsc{$\langle sos \rangle$} at $t=0$ and stopping when the predicted output corresponds to the special end token \textsc{$\langle eos \rangle$} or when the maximum length ($22$ in our experiments) is reached. 

All our captioning models are trained by minimising a Cross Entropy Loss between the probability distribution over the candidate sentence and the ground-truth caption. 

We set the number of hidden units in both the encoder ($H^{enc}$) and the decoder ($H^{dec}$) to 256. As part of the training procedure we use Adam as an optimiser, with a simple learning rate schedule to linearly reduce the learning rate from an initial value of $10^{-4}$. We train for a maximum of 200 epochs with a batch size of $16$, making use of dropout with rate $0.25$ and early stopping with a patience of 10 epochs for regularisation. Training takes roughly one day on a single RTX 2080 Ti GPU.

\begin{table*}[!t] 
\caption{Captioning performance of the ablation models. PT: pre-training; BS: beam search (with beam size $b = 5$ for early fusion, $b = 3$ in all other cases); ATT: attention. We highlight in bold scores within two standard deviations of the best result for each metric.}
\vspace{-0.2cm}
\label{tab:ablations}
\centering
\begin{sc}
\begin{tabular}{lp{0.3cm}p{0.3cm}p{0.4cm}ccccccccc}
\toprule
 Fusion & PT & BS & ATT & BLEU\textsubscript{1} & BLEU\textsubscript{2} &  BLEU\textsubscript{3} &  BLEU\textsubscript{4} & METEOR & ROUGE\textsubscript{L} & CIDEr & SPICE & SPIDEr \\
\midrule
Early &  &  &  & $35.3$ & $14.4$ & $7.6$ 	& $4.6$ 	& $29.3$  & $39.3$ 	& $29.2$ 	& $21.8$  & $25.5$  \\
\cdashlinelr{1-13}
\multirow{2}{*}{Early}
& \checkmark &  &  & $\mathbf{37.8}$ 	& $\mathbf{16.5}$ 	& $\mathbf{9.2}$ 	& $\mathbf{5.8}$ 	& $29.4$ 	& $40.4$ 	& $37.6$ 	& $23.3$ 	& $\mathbf{30.4}$  \\
& \checkmark & \checkmark &  & $33.4$  & $15.0$ 	& $8.7$ 	& $\mathbf{5.7}$ 	& $29.3$ 	& $40.8$ 	& $\mathbf{38.8}$ 	& $23.9$ 	& $\mathbf{31.4}$  \\
\cdashlinelr{1-13}
\multirow{2}{*}{Late}
& \checkmark &  &  & $35.8$  & $15.8$ 	& $9.0$ 	& $\mathbf{5.8}$ 	& $\textbf{29.6}$ 	& $\mathbf{41.0}$ 	& $36.8$ 	& $\mathbf{24.3}$ 	& $30.6$   \\
& \checkmark & \checkmark & &  $32.0$  &	$14.4$  &	$8.5$ 	& $5.6$  &	$29.5$ 	& $\mathbf{41.2}$ 	& $35.9$ 	& $\mathbf{24.7}$ 	& $\mathbf{30.3}$  \\
\cdashlinelr{1-13}
-- & \checkmark & & \checkmark &  $36.7$ 	& $16.2$ 	& $\mathbf{9.2}$ 	& $\mathbf{5.8}$ 	& $\mathbf{29.7}$ 	& $\mathbf{41.0}$ 	& $37.1$ 	& $23.2$ 	& $30.2$  \\ 
-- & \checkmark &  \checkmark & \checkmark  & $33.1$  & $14.7$ 	& $8.6$ 	& $5.6$ 	& $\mathbf{29.5}$  & $\mathbf{41.1}$ 	& $36.7$  & $23.5$ 	& $30.1$   \\
\bottomrule
\end{tabular}
\end{sc}
\vspace{-0.3cm}
\end{table*}

\subsection{Evaluation}
\subsubsection{Baseline}
We compare our method to the publicly available baseline system used for the audio captioning task of the DCASE 2020 Challenge\footnote{https://github.com/audio-captioning/dcase-2020-baseline/}, trained and evaluated on our music captioning dataset. This is a sequence-to-sequence model consisting of three bidirectional GRUs as the encoder, operating on the audio input, and one bidirectional GRU as the decoder, with no alignment mechanism between the two. Since the model is tuned to take as input audio samples between 15s and 30s at a sampling rate of 16 kHz, we randomly select a 30s segment from each of our audio tracks, using the same data splits as in the training and evaluation of our model. To ensure a fair comparison, we empirically verify that training our model with smaller, randomly selected audio chunks does not significantly affect its performance.

\subsubsection{Captioning}
In line with previous work in image and audio captioning, we provide an evaluation of the generated captions using standard automatic metrics. These can be divided into two groups: the first, comprising of BLEU \cite{Papineni2002}, METEOR \cite{Lavie2007} and ROUGE\textsubscript{L} \cite{Lin2004}, is a set of metrics first proposed to evaluate text generated through machine translation; the second, including CIDEr \cite{Vedantam2015}, SPICE \cite{Anderson2016} and SPIDEr \cite{Liu2016}, was instead introduced for the evaluation of image captioning models. BLEU is computed from the geometric mean of the $n$-gram-based precision, using either only unigrams (BLEU\textsubscript{1}), or combinations of $n$-grams up to 4, while METEOR and ROUGE\textsubscript{L} are instead computed from an F-score based on matches between the candidate sentence and the reference one. 
Among the second set of metrics, CIDEr computes the cosine similarity between tf-idf weighted $n$-grams and is observed to better correlate with human judgement compared to previous metrics\cite{Vedantam2015}, SPICE is based on a comparison of semantic similarity of scene graphs parsed from the candidate and reference caption, and SPIDEr is a linear combination of the two. We also note that these metrics have been found to be more accurate when more than one reference sentence is provided \cite{Vedantam2015}, while only one caption per example is available in our dataset. 

\subsubsection{Text-based Music Retrieval}
We also evaluate our model on music audio retrieval based on a natural language query. The goal of this task is to retrieve the correct audio among a pool of candidate tracks, given its corresponding textual description. We note that our model was not trained on retrieval and its objective is not explicitly designed to optimise text-audio ranking, and therefore does not allow for a direct way to retrieve audio given a caption (or vice versa). However, the model can be easily adapted to perform retrieval by ranking each candidate audio track according to the conditional probability of generating the text query given the audio input, corresponding to the joint probability over the $T$ words in the query sentence, as can be seen in \eqref{eq:joint}. For this experiment, we use the 1207 audio tracks in our test set and 100 test queries from the set of ground-truth captions. Following prior work on image description \cite{Vinyals2014, Karpathy2017}, we provide results for the following retrieval metrics: Recall @ $K$ (higher is better) with $k = \{1, 5, 10\}$, the percentage of correctly retrieved items within the top-$K$ results, and the median rank (lower is better) of the ground-truth items across all queries.

\subsection{Results \& Analysis}
In this section, we discuss our experimental results for the captioning and retrieval tasks, comparing several variants of our model.
We also perform an ablation study (Table \ref{tab:ablations}) to tease apart the relative contributions of the main design choices. In order to account for variance due to random initialization, we repeat each experiment in the ablation study three times and provide confidence intervals on our results.
For the captioning task, we look at: (1) comparison with the baseline; (2) the effects of large-scale pre-training of the audio feature extractor module; (3) the choice of modality fusion in the encoder and the role of the attention mechanism in providing temporal alignment between the audio embeddings and the representations learnt by the decoder; (4) additionally, we provide a comparison of model performance and caption quality with two different decoding strategies.

\subsubsection{Comparison with the Baseline}\label{sec:effect_audio_extractor}
In our first experiment, we focus on the comparison between different audio feature extractors (\textit{musicnn} and \textit{Hcnn}), pre-trained on the MagnaTagATune dataset \cite{Law2009} and compare the two resulting variants of our model, denoted as MusCaps-Musicnn and MusCaps-Hcnn. All experiments were run with the same settings, as detailed in Section \ref{sec:settings}, using early fusion and keeping the pre-trained audio feature extractor module frozen during training. For a fair comparison to the baseline, which does not include an attention mechanism, we only include results of our attention-free model in this section.
Our experimental results, shown in Table \ref{tab:results}, indicate that both variants of our model significantly outperform the DCASE baseline across all metrics, suggesting that the use of a musically informed audio feature extractor brings a considerable performance boost. Illustrative examples of the qualitative differences in output captions between our model and the baseline are also reported in Table \ref{tab:examples}.

\begin{figure}[t]
\centering
\includegraphics[scale=0.6]{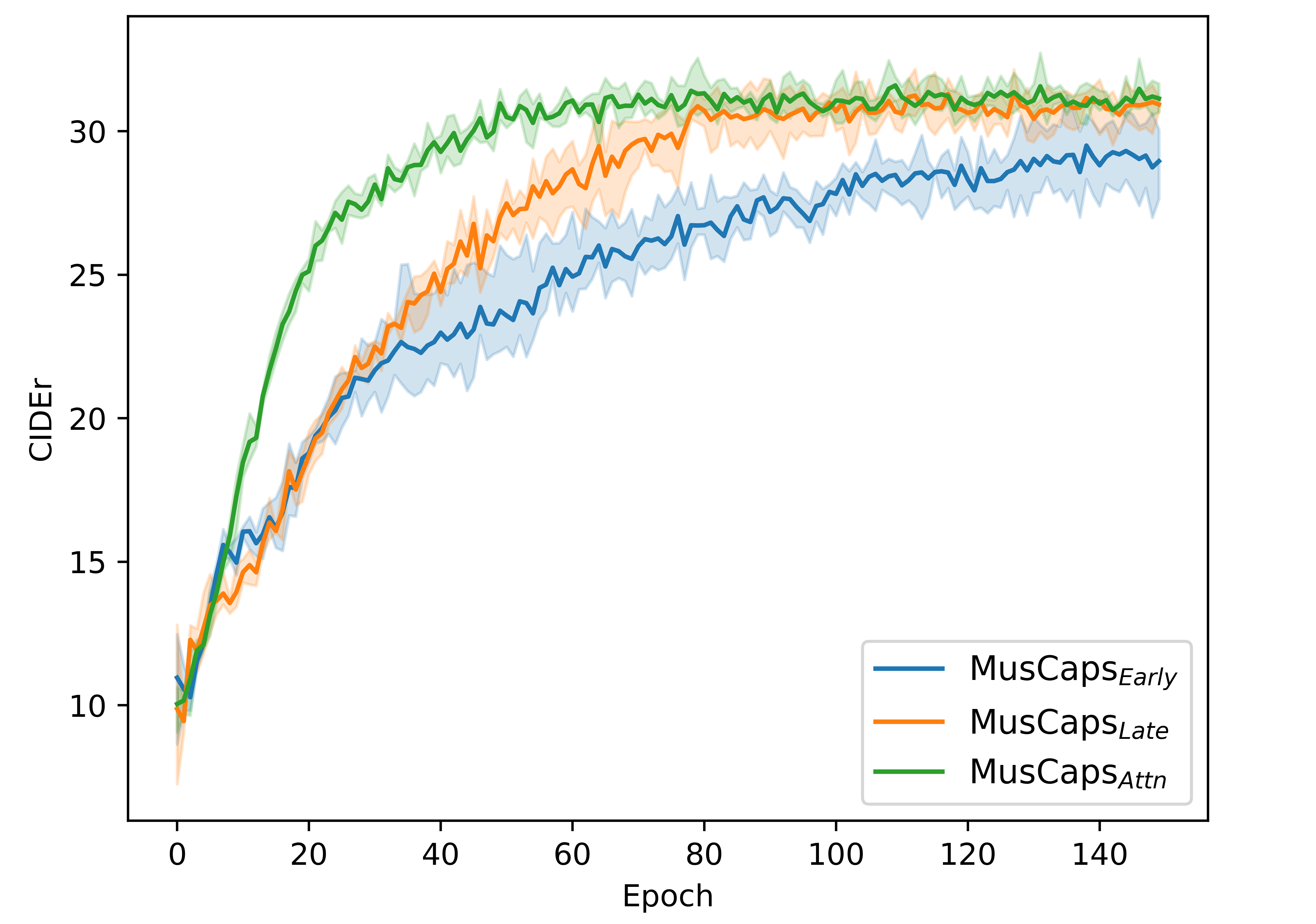}
\vspace{-0.6cm}
\caption{CIDEr performance of the ablation models on the validation set in the first 150 training epochs. Each curve is averaged across 3 runs and the shaded regions indicate 95\% confidence intervals.}
\label{fig:cider}
\vspace{-0.5cm}
\end{figure}

\begin{table*}[!t] 
\caption{Retrieval performance. We highlight in bold scores within two standard deviations of the best result for each metric.}
\vspace{-0.2cm}
\label{tab:retrieval}
\centering
\begin{sc}
\begin{tabular}{lcccc}
\toprule
 Model & Recall@1 $\uparrow$ & Recall@5 $\uparrow$ &  Recall@10 $\uparrow$ &  Median Rank $\downarrow$ \\
\midrule
DCASE Baseline & $1.2$	& $5.2$	& $10.3$ & $41.8$ \\
\cdashlinelr{1-5}
MusCaps w/o pre-training & $1.7$ & $6.0$ & $9.0$	& $ 107.2 $ \\
MusCaps\textsubscript{$Early$} & $2.3$ &	$12.0$ & $21.3$	& $55.2$ \\
MusCaps\textsubscript{$Attn$} & $2.7$ & $\mathbf{15.0}$ & $24.3$ & $45.7$ \\
MusCaps\textsubscript{$Late$} & $\mathbf{5.7}$ & $\mathbf{18.0}$ & $\mathbf{27.3}$ & $\mathbf{37.8}$ \\
\bottomrule
\end{tabular}
\end{sc}
\vspace{-0.5cm}
\end{table*}

\begin{table*}[!t] 
\caption{Examples of predicted captions illustrating the improved sentence quality achieved through beam search (beam size $b = 5$). Cases of repetitions generated through greedy decoding that do not occur when using beam search are highlighted in bold.}
\vspace{-0.2cm}
\label{tab:examples}
\centering
\begin{tabular}{p{2cm}p{3.5cm}p{3.5cm}p{3.5cm}p{3.5cm}}
\toprule
\sc{Model} & \sc{Example 1}  & \sc{Example 2}  & \sc{Example 3}  & \sc{Example 4} \\
\midrule
\sc{Ground Truth}  &  \textit{Relaxed feelgood \textsc{$\langle unk \rangle$} guitars} & \textit{Rousing and emotive epic adventure piece} &  \textit{Dark brooding orchestral motifs with ghostly ethnic woodwind effects} & \textit{Haunting vocal textures over earthy tribal drums and epic orchestra} \\
\cdashlinelr{1-5}
\sc{DCASE Baseline}  &  \textit{Warm and and with acoustic guitar} & \textit{Sweeping and and with strings strings and and} & \textit{Heart-wrenching and and with with} & \textit{Dark and and with to and and and} \\
\cdashlinelr{1-5}
\sc{MusCaps w/ greedy decoding}  & \textit{Upbeat acoustic guitar \textbf{guitar}} & \textit{Powerful and dramatic opening to \textsc{$\langle unk \rangle$} \textsc{$\langle unk \rangle$} and \textsc{$\langle unk \rangle$} \textsc{$\langle unk \rangle$} at \textsc{$\langle unk \rangle$}} & \textit{Dark and \textbf{dramatic} introduction to dramatic orchestral theme} & \textit{Epic \textbf{epic} theme featuring choir and \textbf{choir} and \textsc{$\langle unk \rangle$} \textsc{$\langle unk \rangle$}} \\
\cdashlinelr{1-5}
\sc{MusCaps w/ beam search} & \textit{Upbeat acoustic guitar \textbf{tune}} & \textit{Powerful and ominous introduction builds to dramatic epic theme} & \textit{Dark and \textbf{atmospheric} introduction builds to dramatic theme} & \textit{\textbf{Haunting} introduction with female vocals and haunting strings} \\
\bottomrule
\end{tabular}
\vspace{-0.5cm}
\end{table*}

\subsubsection{Effect of Pre-training}
To further support our claim that pre-training the audio feature extractor module on the music auto-tagging task is indeed crucial to obtaining a good model performance, we empirically verify this by running experiments without initialising the audio encoder parameters with pre-trained weights, training them instead from scratch with the rest of the model.
Due to memory constraints, we randomly select $20s$ chunks from the full track as our audio input. As hypothesised, our experimental results demonstrate that removing pre-training of the CNN component results in a significant performance drop, as shown in Table \ref{tab:ablations}, and slower convergence. This can be attributed to the lack of sufficient training data to successfully accomplish both feature extraction and sentence generation end-to-end, which quickly leads to overfitting, and to a possible difference in the rates at which these two model components generalise. Similarly to what was done when comparing our model with the baseline, we also train MusCaps with pre-training using shorter, randomly selected chunks as the audio input, to ensure that this does not significantly affect performance, and find that only a small decrease in some of the scores is observed in this case.

\subsubsection{Effect of Fusion \& Attention}
We next experiment with the two fusion strategies described in Section \ref{sec:encoder} and the use of a simple attention mechanism for temporal alignment (Section \ref{sec:attention_encoder}), denoting the respective models as MusCaps\textsubscript{$Early$}, MusCaps\textsubscript{$Late$} and MusCaps\textsubscript{$Attn$}. Overall, we do not observe a significant difference across these three variants. Remarkably, we find that, even when including the attention mechanism, the model performs similarly. We identify two main explanations for this. Firstly, the attention mechanism used assumes an alignment between audio segments and word tokens at a consistent temporal scale. In music captioning, however, an exact alignment between discrete units in the audio and text modalities may be too strong of an assumption and, if present, is likely to occur at different timescales.
Secondly, as shown in Fig. \ref{fig:cider}, the learning curves of the ablation models reveal that MusCaps\textsubscript{$Attn$} converges faster than all other variants. This behaviour is more pronounced for CIDEr, but a similar trend is observed on all metrics. This suggests that the model may be overfitting and ultimately warrants further investigation. 

\subsubsection{Effect of Decoding Strategies}
Finally, we compare performance when using greedy decoding and beam search. We find that a beam size of 5 brings the highest performance gain on the captioning metrics for MusCaps\textsubscript{$Early$}, while in most other variants performance saturates quickly and even degrades with a beam size larger than 3. This is consistent with observations from previous studies \cite{Vinyals2017}, which attribute the performance degradation to overfitting or discrepancy between optimisation of the metrics and of the learning objective.

To better understand how caption quality if affected by the use of different decoding algorithms beyond what is captured by automatic metrics, we extract some statistics from the predicted captions.
When greedily decoding the text output, we observe that, although the generated sentences generally capture the audio content well, they often present undesirable features, such as a particularly high occurrence of the \textsc{$\langle unk \rangle$} token and the frequent repetition of $n$-grams (``\textit{and strings and strings}", ``\textit{rock rock}", ``\textit{bass bass and bass}").
When comparing the output of greedy decoding to that obtained through beam search, we observe a significant reduction in both of these. On MusCaps\textsubscript{$Early$}, beam search decoding with $b = 5$ produces $20\%$ fewer \textsc{$\langle unk \rangle$} tokens and $37\%$ fewer repetitions. A similar trend is observed in other architectural variants, such as MusCaps\textsubscript{$Attn$}, although the improvement is more modest ($-11\%$ and $-18\%$ respectively). As shown in Table \ref{tab:examples}, this results in more readable captions and we therefore argue that, although not consistently reflected in the evaluation metrics, beam search improves structure and is therefore beneficial to caption generation overall.

\subsubsection{Retrieval Performance}
In Table \ref{tab:retrieval} we report the audio retrieval performance of the 4 model variants analysed in the ablation study, comparing them to the baseline. The results generally mirror what is observed in the captioning task: pre-training of the audio feature extractor brings the highest performance gain, while changes due to design choices such as fusion strategy and the inclusion of the attention mechanism are less substantial. Surprisingly, unlike in the captioning results, the recall performance of the DCASE baseline model is comparable to MusCaps without pre-training, while its median rank is similar to the top-performing MusCaps variants. 
Overall, although our experiments demonstrate that the model can be used for text-based retrieval, the scores are not particularly high and it remains unclear whether this is due to genuinely incorrect audio-text matching or to limitations of the ranking procedure. It is worth noting, for example, that the task itself may be partially ill-defined in this context, since multiple textual descriptions may be acceptable for a given audio clip (and vice versa), while our dataset, and therefore our evaluation procedure, does not account for this. A more thorough evaluation protocol may include similarity measures between the retrieved and ground-truth items and user studies. 

\section{Conclusions \& Future Work}\label{sec:conclusion}
In this paper, we presented the first music audio captioning model, MusCaps, a simple encoder-decoder network consisting of a multimodal CNN-LSTM encoder with temporal attention  and an LSTM decoder. We formulate the problem in a transfer learning setting and highlight the benefit of leveraging music audio feature extractors pre-trained on large-scale, publicly available datasets. Our experiments show that our model can successfully generate descriptive sentences of a music track and outperforms a general-purpose baseline model for audio captioning, even when training on a small audio-text corpus. This confirms that both a musically informed architecture and the use of music tagging as a pretext task are beneficial to music captioning.
Although our results are encouraging, our work offers several avenues for further study. Among these, data-related limitations warrant particular attention and we emphasise that additional data collection or data augmentation techniques may be required to obtain several reference captions per example and thus make metric-based evaluation more reliable.

\bibliographystyle{IEEEtran}
\bibliography{IEEEabrv, references}

\end{document}